\documentclass{appolb}
\usepackage{graphicx}

\usepackage{amssymb}

\usepackage[thmmarks]{ntheorem-hyper}
\theoremheaderfont{\bf}\theorembodyfont{\it}\theoremsymbol{}\theoremstyle{plain}\newtheorem{Thm}{Theorem}
\theoremheaderfont{\bf}\theorembodyfont{\rm}\theoremsymbol{\setlength{\unitlength}{1mm}\begin{picture}(2.5,2.5)\linethickness{2.5mm}\put(0,1.25){\line(1,0){2.5}}\end{picture}}\theoremstyle{plain}
\theoremheaderfont{\bf}\theorembodyfont{\it}\theoremsymbol{}\theoremstyle{plain}
\theoremheaderfont{\bf}\theorembodyfont{\sl}\theoremsymbol{}\theoremstyle{plain}\newtheorem{Rem}[Thm]{Remark}
\theoremheaderfont{\bf}\theorembodyfont{\it}\theoremsymbol{}\theoremstyle{plain}
\theoremheaderfont{\bf}\theorembodyfont{\sl}\theoremsymbol{}\theoremstyle{plain}
\theoremheaderfont{\bf}\theorembodyfont{\it}\theoremsymbol{}\theoremstyle{plain}

\newcommand{\R}{\mathbb{R}}

\newcommand{\N}{\mathbb{N}}
\newcommand{\LL}{\mathbf{L}}
\newcommand{\vv}{\mathbf{v}}
\newcommand{\Real}{\mathop{\mathrm{Re}}\nolimits}
\newcommand{\Imag}{\mathop{\mathrm{Im}}\nolimits}
\newcommand{\I}{\mathrm{i}}
\newcommand{\Q}{\mathcal{Q}}
\newcommand{\D}{\mathrm{D}}
\newcommand{\de}{\mathrm{d}}

\begin{document}
\title{A hidden symmetry of conformally invariant Lagrangians}
\author{Andr\'as L\'aszl\'o
\address{Wigner Research Centre for Physics\\
Konkoly-Thege M.\ u.\ 29-33., 1121 Budapest, Hungary\\
\texttt{laszlo.andras@wigner.hu}}
}
\maketitle
\begin{abstract}
In this paper a hidden extra symmetry of conformally invariant Lagrangians 
occuring in physics is pointed out. This symmetry is most apparent in 
a metric independent, i.e.\ in a Palatini-like presentation 
of the variational problem. 
In such presentation, the usual conformal weight of fields 
can be encoded as local dilatation group gauge charges. The conventional 
conformal invariance of Lagrangians is then equivalent to dilatation gauge 
invariance. The claim of the paper is, that the most 
commonly occurring conformally invariant Lagrangians turning up in 
physics are not only invariant to local dilatation gauge transformations, 
but they are also invariant to any change of the dilatation gauge connection, 
meaning an additional algebraic symmetry property. In terms of dimensional 
analysis and differential geometry, this additional symmetry means complete 
insensitivity of the Lagrangian to the choice of the parallel transport 
rule of local measurement units throughout spacetime.
\end{abstract}

\section{Introduction}

Conformally invariant field theories form a very important class of models 
when it comes to general relativistic (GR) model building related to particle 
physics. For instance, the kinetic terms of the general relativistically 
formulated Standard Model (SM) Lagrangian are all conformally invariant, or 
can be made conformally invariant with rather plausible generalization.

Given a general relativistic spacetime model $(M,g)$, $M$ being a four 
dimensional real smooth manifold and $g$ being a smooth Lorentz signature 
metric tensor field over it, a conformal transformation is a pair 
$(\phi,\Omega)$, where $\phi$ is an $M\rightarrow M$ diffeomorphism and 
$\Omega$ is an $M\rightarrow\R^{+}$ positive valued smooth scalar field. 
A conformal transformation $(\phi,\Omega)$ maps a metric tensor field $g$ 
to an other one by the group action $\Omega^{2}\phi^{*}g$, where 
$\phi^{*}$ denotes the pullback operation of the diffeomorphism 
$\phi$.\footnote{The group of conformal transformations is not to be confused with 
the conformal diffeomorphism group, which is a subgroup of the group of 
conformal transformations, satisfying the condition $\Omega^{2}\phi^{*}g=g$, also 
called as conformal isometries.}
An important subgroup of these transformations are the 
conformal rescalings, also called \emph{Weyl rescalings}. For Weyl rescalings, 
the diffeomorphism $\phi$ is the identity of $M$ and $\Omega$ is kept to be an arbitrary positive 
valued smooth scalar field. In a field theory, a group action of the Weyl rescalings 
on the fundamental fields may be specified, called to be the conformal weights, 
and then the conformal group can act on all the fields in the theory simultaneously. 
Whenever the field equations or the action functional of the model is invariant to 
such a group action, the theory is called conformally invariant 
\cite{wald1984,penrose1984,hawking1973}.

In the above, conventional definition of conformal invariance one needs 
first a field equation or Lagrangian, and the spacetime metric must be one of 
the fundamental fields. Then, a group action of the Weyl rescalings of the metric 
on all the fields must be specified. Only if all these group actions are declared, then 
the invariance of the field equations or the action functional can be stated.

In this paper, a less metric dependent formulation of conformal 
invariance is used. In that approach, all the fundamental fields are regarded as fields 
with $\D(1)$ gauge charge, analogous to the conformal weight.\footnote{The 
\emph{dilatation group}, $\D(1)$, is $\R^{+}$ with the real multiplication.} 
The spacetime metric tensor field may as well be a composite field, i.e.\ some 
function of more fundamental fields, such as its spinorial decomposition. 
In such $\D(1)$ gauge theory reformulation, 
the usual conformal invariance in terms of Weyl rescalings is equivalent to a requirement on 
the action functional to be $\D(1)$ gauge invariant. 
The claim of the paper is that the most common conformally invariant 
Lagrangians in physics have a slightly larger symmetry than mere conformal 
invariance: their action functional has an algebraic property of being not 
only $\D(1)$ gauge invariant, but to be completely invariant to the choice of 
the $\D(1)$ gauge connection.

The mentioned phenomenon can most easily be seen on a general relativistic 
Dirac Lagrangian $\vv_{\gamma}\Real\left(\overline{\Psi}\gamma^{a}\I\nabla_{a}\Psi\right)$, 
where $\gamma_{a}$ is the Clifford map associated with a Lorentz metric $g_{ab}$, 
$\vv_{\gamma}$ is the volume form subordinate to the Clifford map (or equivalently, 
to the metric), $\Psi$ is a Dirac field, and $\nabla_{b}$ is the combined 
metric and $\mathrm{U}(1)$ gauge covariant derivation \cite{Trautman2008}. (Penrose abstract 
indices are used for the tangent indices.) This Lagrangian is very well known 
to be conformally invariant, i.e.\ invariant to the transformation
$\big(\Psi,\,\gamma_{a},\,\nabla_{b}\big)\mapsto\big(\Omega^{{}_{-\frac{3}{2}}}\Psi,\,\Omega\gamma_{a},\,\nabla_{b}-\frac{1}{2}(\I\Sigma_{b}{}^{c}{-}\delta_{b}{}^{c}I)(\Omega^{{}_{-1}}\de_{c}\Omega)\big)$, 
where $\Sigma_{ab}:=\frac{\I}{2}\left(\gamma_{a}\gamma_{b}-\gamma_{b}\gamma_{a}\right)$ 
stands for the \emph{spin tensor}. The transformation rule of $\nabla_{b}$ 
is very well known to be uniquely determined by the requirement that the 
transformation preserves the metricity, torsion-freeness and the compatibility 
to the Clifford map. This ordinary Dirac 
Lagrangian may be generalized in a quite straightforward way: the covariant 
derivation $\nabla_{b}$ may be generalized by incorporating a $\D(1)$ gauge 
potential as well. In that case, the pertinent Dirac Lagrangian is 
invariant as well to a slightly different transformation 
$\big(\Psi,\,\gamma_{a},\,\nabla_{b}\big)\mapsto\big(\Omega^{{}_{-\frac{3}{2}}}\Psi,\,\Omega\gamma_{a},\,\nabla_{b}+\Omega^{{}_{-\frac{3}{2}}}\de_{b}\Omega^{{}_{\frac{3}{2}}}\big)$, 
which can be considered as a local $\D(1)$ gauge transformation. 
With this generalization, the gradients of $\Omega$ are absorbed by the $\D(1)$ 
gauge potential, understood within $\nabla_{b}$. In such variables, it 
is straightforward to verify that the Dirac Lagrangian is invariant to a 
further transformation: 
$\big(\Psi,\,\gamma_{a},\,\nabla_{b}\big)\mapsto\big(\Psi,\,\gamma_{a},\,\nabla_{b}+C_{b}\big)$, 
where $C_{b}$ is any $\D(1)$ gauge potential, i.e.\ is an arbitrary real valued 
smooth covector field.

In this paper, we show that the pertinent shift symmetry
\begin{eqnarray}
 \nabla_{b} & \mapsto & \nabla_{b} + C_{b} \qquad \mathrm{(} C_{b} \mathrm{\;being\;a\;} \D(1) \mathrm{\;gauge\;potential)}
\end{eqnarray}
is also the symmetry of any conformally invariant Lagrangian turning up 
in physics.
This algebraic symmetry property is in addition to their conformal invariance, 
and it means that the pertinent type of Lagrangians are completely insensitive 
to the parallel transport rule of local measurement units throughout spacetime. 
Although conformal invariance has been studied extensively 
(see a comprehensive review in \cite{Iorio1996,Iorio1997}), to our knowledge such an 
algebraic symmetry property has not yet been explicitly pointed out in the literature.

The structure of the paper is as follows. First, we recall the Lagrangian formulation of classical 
field theories in metric independent way, i.e.\ in a Palatini-like approach. 
Then, we invoke a simple formalism for endowing the fields with 
$\D(1)$ charges without referring to a metric. 
Following that, we shall present important examples of conformally invariant 
Lagrangians and show that they have the additional algebraic symmetry property of being invariant 
to the choice of the $\D(1)$ connection, which is the emphasis of the paper. 
Then, we show a (non-physical) quite trivial counterexample of a 
conformally invariant Lagrangian, which does not have the pertinent additional 
symmetry property. Finally, we conclude.

\section{Non-metric formulation of classical field theories}
\label{sectioncft}

In this section the precise mathematical definition of classical field 
theories is recalled in terms of the Lagrangian and variational principles: 
for a comprehensive overview, see e.g.\ \cite{laszlo2004,pons2003,peldan1994}. 
The used definition deliberately does not refer to an a priori 
known spacetime metric tensor field, and thus resembles basically to a 
Palatini type formulation \cite{wald1984}. In the followings, we shall 
denote the tangent bundle of a manifold $M$ by $T(M)$, and by $T^{*}(M)$ 
the corresponding cotangent bundle. The vector bundle of maximal forms 
is denoted by ${\mathop{\wedge}\limits^{n}}T^{*}(M)$ 
where $n:=\dim(M)$. In the followings every differential geometrical object is 
assumed to be smooth for simplicity of presentation: strict differentiability 
counting is performed in \cite{laszlo2004}. The vector space of smooth sections of 
some vector bundle $V(M)$ over $M$ is denoted by $\Gamma(V(M))$. 
The affine space of covariant derivations over $V(M)$ shall be denoted by 
$D(V(M))$. The corresponding dual vector bundle of $V(M)$ is denoted by 
$V^{*}(M)$. These notations are the usual ones in differential geometry 
literature. In addition, we shall use Penrose abstract indices 
\cite{penrose1984,wald1984} for denoting tensor traces and 
expressions concerning the tensor powers of $T(M)$ and $T^{*}(M)$. The 
abstract indices of $T(M)$ shall be denoted by superscripted lower case latin 
letters ($\,{}^{abcd\dots}$), whereas for $T^{*}(M)$ subscripted lower case 
latin letters ($\,{}_{abcd\dots}$) shall be used. The index symmetrization 
operation shall be denoted by curly brackets, e.g.\ $t_{(abc)}$, whereas 
the antisymmetrization operation shall be denoted by square brackets, 
e.g.\ $t_{[abc]}$, furthermore their normalization convention shall be set 
as in e.g.\ \cite{wald1984}.

Recall that the space of smooth sections $\Gamma(V(M))$ of some vector bundle 
$V(M)$ admits a natural $\mathcal{E}$ test function topology \cite{laszlo2004}: 
without any further assumption it is meaningful to define convergence of a sequence 
$(\varphi_{k})_{k\in\N}$ in $\Gamma(V(M))$ to a limit $\varphi$ in 
$\Gamma(V(M))$ with requiring that the field $(\varphi-\varphi_{k})_{k\in\N}$ and 
all of its gradients uniformly converge to zero on any compact region of 
$M$.\footnote{A pointwise change of the norms and covariant derivation operators 
acting on $\Gamma(V(M))$, used for the definition of such convergence notion, form a norm equivalence class in each point of $M$, 
as e.g.\ summarized in \cite{laszlo2013}~Appendix~A~Lemma~3. Because of that pointwise 
norm equivalence, the $\mathcal{E}$ convergence notion does not depend on 
the particular choice of these auxiliary mathematical objects.}
Whenever the manifold $M$ is compact, or a fixed compact region 
$K\subset M$ is considered, the $\mathcal{E}$ topology naturally gives rise to 
a norm equivalence class on the fields over the pertinent region \cite{laszlo2004,laszlo2013}. 
Because of that, ordinary (Fr\'echet) derivatives of functionals of such local fields 
can be naturally defined without further mathematical assumptions.

As usual in the differential geometry literature \cite{wald1984}, a covariant derivation 
on a vector bundle $V(M)$ may be uniquely extended to all the 
tensor powers of $V(M)$ and its dual bundle $V^{*}(M)$ by requiring Leibniz rule over tensor product, 
commutativity with tensor contraction, and correspondence to the exterior derivation 
over the scalar line bundle $M\times\R$. Similarly, given two different vector 
bundles over $M$ along with covariant derivation on each, then they naturally give 
rise to a joint covariant derivation, which uniquely extends 
to all tensor powers of the pertinent vector bundles and their duals, by 
requiring analogous properties.

\vspace*{2mm}
\begin{Rem}
If $\nabla$ is a covariant derivation over $T(M)$, then there is a unique 
covariant derivation $\tilde{\nabla}$ over $T(M)$ associated to it, having 
vanishing torsion tensor and having the same geodesics as $\nabla$. The covariant 
derivation $\tilde{\nabla}$ is called the \emph{torsion-free part} of $\nabla$. 
In explicit formulae: whenever $v^{b}$ is a smooth section of $T(M)$, then one has
$\tilde{\nabla}_{a}v^{b}=\nabla_{a}v^{b}+\frac{1}{2}T(\nabla)_{ac}^{b}v^{c}$, where 
$T(\nabla)_{ac}^{b}$ denotes the torsion tensor of $\nabla$.
\label{torsionremark}
\end{Rem}

\vspace*{2mm}
\begin{Rem}
Let $J^{a}_{[c_{1}\dots c_{n}]}$ be a smooth section of 
$T(M)\otimes \mathop{\wedge}\limits^{n}T^{*}(M)$, i.e.\ a maximal form valued tangent vector 
field. Then, given any covariant derivation $\nabla$ on $T(M)$, 
one has that the expression 
$\tilde{\nabla}_{a}J^{a}_{[c_{1}\dots c_{n}]}$ is independent of the choice of 
the covariant derivation, where $\tilde{\nabla}$ denotes the torsion-free part 
of $\nabla$. That is, the divergence of a maximal form valued vector field is 
naturally defined without further assumptions. Similarly, for a smooth section 
$K^{[ab]}_{[c_{1}\dots c_{n}]}$ of $T(M){\wedge}T(M)\otimes \mathop{\wedge}\limits^{n}T^{*}(M)$ 
one has that $\tilde{\nabla}_{a}K^{[ab]}_{[c_{1}\dots c_{n}]}$ is independent 
of the choice of the covariant derivation and thus the divergence of such field 
is naturally defined without further assumptions.
\label{divergenceremark}
\end{Rem}
\vspace*{2mm}

Given the above notions and observations, a classical field theory may be defined as a quartet 
\begin{eqnarray}
\left(M,V(M),\LL,S\right),
\end{eqnarray}
where $M$ is some finite dimensional differentiable manifold 
possibly with boundary (this is called the \emph{base manifold} --- it models 
the spacetime or a compactified spacetime with or without a boundary), 
$V(M)$ is some finite dimensional smooth vector bundle over it, 
called the \emph{vector bundle of matter fields}. 
The \emph{Lagrange form} $\LL$ is then a smooth pointwise fiber bundle morphism
\begin{eqnarray}
 V(M)\,\times\, T^{*}(M){\otimes}V(M)\,\times\,T^{*}(M){\wedge}T^{*}(M){\otimes}V(M){\otimes}V^{*}(M) \cr
 \qquad\qquad\qquad\qquad\qquad\qquad\qquad\qquad\qquad\rightarrow \mathop{\wedge}\limits^{n}T^{*}(M),
\label{dLhomom}
\end{eqnarray}
taking the triplet of matter fields, matter field gradients, and field strength tensors 
into a maximal form field. In particular, it acts on the sections as
\begin{eqnarray}
 \LL:\cr
 \; \Gamma\left(V(M)\,\times\, T^{*}(M){\otimes}V(M)\,\times\,T^{*}(M){\wedge}T^{*}(M){\otimes}V(M){\otimes}V^{*}(M)\right) \cr
 \qquad\qquad\qquad\qquad\qquad\qquad\qquad\qquad\qquad\rightarrow \Gamma\left(\mathop{\wedge}\limits^{n}T^{*}(M)\right), \cr
 \;(v,Dv,F)\mapsto \LL(v,Dv,F).
\end{eqnarray}
A pair $(v,\nabla)\in \Gamma(V(M))\times D(V(M))$ is called a \emph{field configuration}, 
which form an affine space over the vector space \emph{field variations} 
$\Gamma(V(M))\times \Gamma(T^{*}(M){\otimes}V(M){\otimes}V^{*}(M))$. 
Given a field configuration $(v,\nabla)$, the map $(v,\nabla)\mapsto\LL(v,\nabla v, F(\nabla))$ is called the \emph{Lagrangian expression}, 
where $\nabla v$ is the covariant derivative of $v$ by $\nabla$, 
and where $F(\nabla)$ denotes the curvature tensor of $\nabla$. 
Then, the \emph{action functional} $S(K)$ is defined on a compact region $K\subset M$ as the integral of the 
Lagrangian expression over $K$:
\begin{eqnarray}
 S(K):\cr
 \quad\Gamma(V(M)) \times D(V(M)) \rightarrow \R,\cr
 \quad(v,\nabla)\mapsto S_{v,\nabla}(K):=\int_{K}\LL(v,\nabla v, F(\nabla)).
\end{eqnarray}
(As such, the action functional can be regarded as a Radon measure valued map 
$S:\,(v,\nabla)\mapsto S_{(v,\nabla)}(\cdot)$ from the field configurations.) 
As usually, the solutions of the field equation of the field theory shall be 
the stationary points of the action functional with the fields having 
fixed boundary value on $\partial K$. More concretely, the field 
$(v,\nabla)\in\Gamma(V(M))\times D(V(M))$ is said to be a solution of the 
field theory whenever for all compact regions $K\subset M$ one has
\begin{eqnarray}
D^{\circ}S_{v,\nabla}(K)=0,
\label{sinnerderiv}
\end{eqnarray}
where $D^{\circ}S(K)$ denotes the Fr\'echet derivative $DS(K)$ of $S(K)$ 
restricted in its linear variable to the space of vanishing field variations on the boundary set $\partial K$.
In the end, as quite expected \cite{laszlo2004}, this is equivalent to the Euler-Lagrange 
equations
\begin{eqnarray}
\label{elin}
 \Big. D_{1}\LL(v,\nabla v,F(\nabla)) 
  -\tilde{\nabla}_{a}D_{2}^{a}\LL(v,\nabla v,F(\nabla)) & = & 0,\cr
 \Bigg. D_{2}\LL(v,\nabla v,F(\nabla))(\cdot)v 
  -\tilde{\nabla}_{a}2D_{3}^{[ab]}\LL(v,\nabla v,F(\nabla))(\cdot) & = & 0
\end{eqnarray}
for the fields $(v,\nabla)$ throughout the interior of any 
compact region $K\subset M$ and thus throughout $M$. 
Here $D_{1}\LL$, $D_{2}\LL$, $D_{3}\LL$ mean the spacetime pointwise partial derivative 
of $\LL$ with respect to its first, second and third argument, respectively, 
i.e.\ the derivative of the Lagrange form along the matter fields, the 
matter field gradients, and the curvature tensor. One should note that 
because of Remark~\ref{divergenceremark}, the covariant derivation 
may be chosen arbitrarily over $T(M)$ in the divergence expressions of 
Eq.(\ref{elin}).

\vspace*{2mm}
\begin{Rem}
Note that whenever a model is considered in which $M$ is compact 
(possibly with boundary), then the field equations can be written in a 
simpler form 
\begin{eqnarray}
\label{sderiv}
DS_{v,\nabla}(M)=0.
\end{eqnarray}
This is quite similar to as in Eq.(\ref{sinnerderiv}), but variation on the 
boundary does not need to be excluded. If the variation on the boundary is 
not suppressed, then along with the Euler-Lagrange equations Eq.(\ref{elin}), 
one gets additional boundary field equations
\begin{eqnarray}
\label{elboundary}
D_{2}^{a}\LL(v,\nabla v,F(\nabla)) & = & 0,\cr
2D_{3}^{[ab]}\LL(v,\nabla v,F(\nabla))(\cdot) & = & 0
\end{eqnarray}
over $\partial{M}$, which can eventually be used to impose boundary 
constraints on the fields.
\label{remarkelboundary}
\end{Rem}
\vspace*{2mm}

For clarity, we note that in the standard GR terminology the above 
approach resembles to the Palatini action principle: the covariant derivation 
is varied independently from the field quantities, in particular, 
independently from the metric tensor field.

\section{Non-metric formulation of conformal invariance}
\label{sectionconfinv}

Given a metric independent formulation of a field theory 
$\left(M,V(M),\LL,S\right)$ as in the previous section, we introduce a metric 
independent notion of conformal weights. For this, we assume that the vector 
bundle of fields $V(M)$ is composed of sectors having \emph{$\D(1)$ gauge charges}, i.e.\ 
\begin{eqnarray}
  V(M)=\mathop{\oplus}\limits_{q\in\Q}V_{q}(M),
\label{vmdef}
\end{eqnarray}
where $\Q$ is a finite set of rational numbers and a $\D(1)$ 
gauge transformation is represented by a nonvanishing smooth field 
$\Omega:M\rightarrow\R^{+}$ acting as
\begin{eqnarray}
 v_{q}  & \mapsto & \Omega^{q}\,v_{q}           \qquad\qquad\qquad\qquad\qquad\quad\,  (v_{q}\in V_{q}(M),\;q\in\Q),\cr
 \nabla & \mapsto & \Omega^{q}\nabla\Omega^{-q}=\nabla-q\,\mathrm{d}\left(\ln\Omega\right) \qquad (\mathrm{over\;}V_{q}(M),\;q\in\Q)
\label{d1gauge}
\end{eqnarray}
on the fields and covariant derivations, where $\mathrm{d}$ denotes exterior derivation. 
The numbers $q\in\Q$ are called \emph{$\D(1)$ gauge charges}, and such transformations are called \emph{$\D(1)$ gauge transformations}.
Whenever a spacetime metric tensor field is present in the theory with nonzero 
$\D(1)$ gauge charge, then quite evidently, the 
field rescalings induced by the $\D(1)$ group can always be redefined such that the 
spacetime metric tensor field has $\D(1)$ gauge charge $2$ by 
convention, i.e.\ belonging to $V_{2}(M)$, which is just an equivalent 
reformulation of the Weyl rescaling, i.e.\ that the metric transforms as 
$g_{ab}\mapsto\Omega^{2}g_{ab}$ by convention. 
With these notions, a field theory $\left(M,V(M),\LL,S\right)$ 
is said to be \emph{$\D(1)$ gauge invariant} whenever its action 
functional is invariant to the $\D(1)$ gauge transformations as in Eq.(\ref{d1gauge}).

It shall be shown in the following sections that the conformally invariant 
Lagrangians turning up in physics have a slightly larger symmetry than 
simple $\D(1)$ gauge invariance: their action functional does not 
depend on the $\D(1)$ gauge connection at all. In terms of 
formulas, this means an algebraic property of the invariance of the 
Lagrangian expression $\LL(v,\nabla{v},F(\nabla))$ to the transformation
\begin{eqnarray}
  v      & \mapsto &  v,\cr
  \nabla & \mapsto & \nabla + q\,C \qquad\mathrm{(over\;each\;sector\;}V_{q}(M)\mathrm{,\;}q\in\Q\mathrm{)}
\label{transf}
\end{eqnarray}
for any real smooth covector field $C\in\Gamma\left(T^{*}(M)\right)$. 
This property shall be called \emph{$\D(1)$ connection invariance}, and is 
seen to be a further symmetry on top of invariance to simple Weyl rescalings. 
It means that the theory is invariant to the choice of the parallel transport 
of local measurement units throughout spacetime.

\subsection{\bf More geometric reformulation using \emph{measure line bundles}}
\label{subsectionmeasurelines}

The notion of $\D(1)$ gauge charge can be reformulated in a 
geometrically even more elegant setting. The key idea is motivated by a work 
of Matolcsi \cite{mato1993} and of Jany\u{s}ka, Modugno, Vitolo \cite{modugno2010}, 
in which they proposed a simple mathematical framework for formal mathematical 
handling of physical units. 
In their concept, the mathematical model of special relativistic 
spacetime is considered to be a triplet $(M,L,\eta)$, where $M$ is a four 
dimensional real affine space (modeling the flat spacetime), $L$ is a one 
dimensional vector space (modeling the one dimensional vector space of 
length values), 
and $\eta:\mathop{\vee}\limits^{2}T\rightarrow \mathop{\otimes}\limits^{2}L$ 
is the flat Lorentz signature metric (constant throughout the spacetime), where 
$T$ is the underlying vector space of $M$ (``tangent space''). The important idea in that construction is: the 
field quantities, such as the metric tensor, are not simply 
real valued, but they take their values in the rational tensor powers of the 
\emph{measure line} $L$.\footnote{The term 
\emph{measure line} was introduced by \cite{mato1993}, whereas the same 
concept is called \emph{scale space} by \cite{modugno2010}. Apparently, 
these two group of authors discovered the pertinent rather useful notion independently.} 
Such a setting formalizes the physical expectation 
that quantities actually have physical dimensions (the metric carries 
length-square dimension in this case), and that quantities with different physical 
dimensions cannot be added since they reside in different vector spaces. 
It is seen that the technique of measure lines is nothing but 
the precise mathematical formulation of dimensional analysis.

Such mathematically precise formulation of dimensional analysis, although 
may seem to be a relatively innocent idea at a first glance, becomes quite powerful 
tool when carried over to a general relativistic framework. 
Namely, let our base manifold $M$ be some four dimensional real manifold 
(with or without boundary), and let $L(M)$ be a real vector bundle over $M$, 
with one dimensional fiber. The fiber of $L(M)$ over each point of $M$ shall model 
the vector space of length values, and the pertinent line bundle shall be called 
the \emph{measure line bundle}, or \emph{line bundle of lengths}. Just like proposed in \cite{mato1993,modugno2010}, the 
field quantities shall carry certain tensor powers of $L(M)$. 
For simplicity, the notation $L^{n}(M):=\mathop{\otimes}\limits^{n}L(M)$ and 
$L^{-n}(M):=\mathop{\otimes}\limits^{n}L^{*}(M)$ shall be used, for all non-negative integers $n$, 
conforming to the conventions of \cite{mato1993,modugno2010}, and also to our physical intuition 
of dimensional analysis.\footnote{Note, that because of the one dimensionality 
of the measure line $L$, rational tensor powers are also well defined as seen in \cite{mato1993,modugno2010}. 
Roughly speaking, if $L$ is a one dimensional real vector space, then $\sqrt[n]{L}$ is 
defined to be the one dimensional vector space obeying $\mathop{\otimes}\limits^{n}\left(\sqrt[n]{L}\right)\equiv L$ for 
a given non-negative integer $n$.} 
The proposed idea can be physically formulated as: the field 
quantities are tagged with physical dimensions, 
but the units of the physical dimensions in different spacetime points are not 
necessarily comparable a priori, but a connection over $L(M)$ needs 
to be explicitly specified for that. Using these conventions, and the analogy of 
Eq.(\ref{vmdef}) we assume that the 
structure of the vector bundle of matter fields takes the form
\begin{eqnarray}
  V(M)        & = & \mathop{\oplus}_{q\in\Q} V_{q}(M), \cr
 & & \cr
  \mathrm{with\quad} V_{q}(M) & = & L^{q}(M) \otimes \mathcal{V}_{q}(M) \qquad \mathrm{(for\;all\;}q\in\Q\mathrm{)}.
\label{vml}
\end{eqnarray}
This form of Eq.(\ref{vmdef}) helps to book-keep the physical dimensions of quantities in a quite transparent way: 
the rational numbers $q\in\Q$ are called \emph{physical dimensions}, the factors 
$L^{q}(M)$ are seen to count the physical dimensions, and $\mathcal{V}_{q}(M)$ would represent 
the dimension-free form of field quantities, but the true physical fields reside 
in $V_{q}(M) = L^{q}(M) \otimes \mathcal{V}_{q}(M)$, carrying appropriate dimensions. 
The pointwise $L(M)\rightarrow L(M)$ vector bundle automorphisms are equivalent to 
$\D(1)$ gauge transformations as discussed previously, and the power 
$q$ in the $L^{q}(M)$ factor shall then automatically correspond to the 
$\D(1)$ gauge charge. Thus, using the formalism of measure line bundle 
makes the $\D(1)$ gauge charge, i.e.\ the physical dimension of the 
fields explicit. From the dimensional analysis point of view all this can simply 
be understood as: the fields are tagged by physical dimensions, but the 
unit of measurement might be spacetime point dependent. 
Since only pure real valued maximal form fields may be integrated throughout the 
manifold, $\LL$ must be pure $\mathop{\wedge}\limits^{n}T^{*}(M)$ valued, without 
physical dimension in terms of powers of $L(M)$. This consistency requirement 
already poses algebraic constraints on the possible Lagrangians, and shows the advantage 
of not neglecting the physical dimensions of fields in the formalism.

It is evidently seen that the property of $\D(1)$ connection 
invariance of an above type model is just equivalent to the independence of the Lagrangian expression 
$\LL(v,\nabla{v},F(\nabla))$ from the $L(M)$ connection. We shall call such models \emph{measure line 
connection invariant}. In the coming section it is shown that the conformally 
invariant Lagrangians turning up in physics possess this property, which means 
that these type of Lagrangians are insensitive to the parallel transport rule 
of measurement units throughout spacetime.

\section{Examples}
\label{examples}

In the followings important conformally invariant Lagrangians are recalled. 
It is shown that these all have an additional symmetry of being invariant to the choice 
of the connection on the measure line bundle, or equivalently, to the choice of 
the $\D(1)$ gauge connection.

\subsection{\bf Conformal invariant version of vacuum general relativity}
\label{vacuumgr}

For illustrative purpose, we present the formulation of the conformally 
invariant generalization of vacuum GR. The model is specified via 
a slightly generalized form of the Einstein-Hilbert Lagrangian. 
Namely, let the base manifold $M$ be $4$ real dimensional and oriented, 
and the vector bundle of fields to be 
$V(M):=L^{-1}(M)\,\oplus\, L^{2}(M){\otimes}\mathop{\vee}\limits^{2} T^{*}(M)$, 
where $L(M)$ is the line bundle of lengths. Let the symbol $\vv(g)$ denote 
the canonical volume form field generated by the dimensional metric tensor field $g\in\Gamma\Big(L^{2}(M){\otimes}\mathop{\vee}\limits^{2}T^{*}(M)\Big)$, 
taking its values in $\Gamma\Big(L^{4}(M){\otimes}\mathop{\wedge}\limits^{4}T^{*}(M)\Big)$, i.e.\ having 
dimension length to the four, as physically expected. We take then the Lagrange 
form to be
\begin{eqnarray}
\label{eh}
\LL: \cr
 \quad \Gamma\left(V(M)\,{\times}\,T^{*}(M){\otimes}V(M)\,{\times}\,T^{*}(M){\wedge}T^{*}(M){\otimes}V(M){\otimes}V^{*}(M)\right) \cr
 \qquad\qquad\qquad\qquad\qquad\qquad\qquad\qquad\quad\rightarrow \Gamma\left(\mathop{\wedge}\limits^{4} T^{*}(M)\right),\cr
 \quad ((\varphi,g_{ab}),\,({D\varphi}_{c},{D{g_{de}}}_{f}),\,(r_{gh},{R_{ghi}}^{j}))\;\mapsto\;\vv(g)\varphi^{2}g^{km}\delta^{l}{}_{n}{R_{klm}}^{n}.
\end{eqnarray}
As already mentioned in Section~\ref{sectioncft}, 
in our variational scheme the quantities are varied independently, i.e.\ no 
a priori relation is assumed between the metric and covariant derivation, 
furthermore, also the torsion of the covariant derivation is not restricted 
initially. It is seen that Eq.(\ref{eh}) simply corresponds to the standard 
Einstein-Hilbert Lagrangian with a slight generalization: the inverse Planck 
length (here denoted by $\varphi$) is not assumed to be constant, but 
can (must) have location dependence, i.e.\ it is rather a field than a 
constant in this model, as it is set to be a section of the vector bundle $L^{-1}(M)$. 
With this simple generalization, the theory becomes measure line connection 
invariant, and hence $\D(1)$ connection invariant in terms of our definition 
in Section~\ref{sectionconfinv}. This is verified by directly observing that 
at any field configuration $\left((\varphi,g_{ab}),\nabla_{c}\right)$ the Lagrangian expression 
\begin{eqnarray}
 \vv(g)\varphi^{2}g^{ab}{R(\nabla)_{acb}}^{c}
\end{eqnarray}
is invariant to the transformation Eq.(\ref{transf}), i.e.\ does 
not depend on the covariant derivation over the line bundle of lengths, where 
$R(\nabla)_{acb}{}^{d}$ is the Riemann tensor of $\nabla$.

The field equations are derived by direct substitution of $\LL$ in 
Eq.(\ref{eh}) into Eq.(\ref{elin}), along with subsequent usage of the 
identities $\frac{\partial{\vv(g)}}{\partial{g_{ab}}}=\frac{1}{2}g^{ab}\vv(g)$ 
and $\frac{\partial{g^{cd}}}{\partial{g_{ab}}}=-\frac{1}{2}\left(g^{ca}g^{bd}+g^{cb}g^{ad}\right)$. 
Straightforward calculations show (see also \cite{laszlo2004}), that the field equations read as
\begin{eqnarray}
\label{eprelim}
 \Big. \tilde{\nabla}_{a}(\varphi^{2}g_{bc})  & = & 0,\cr
 \Big. \varphi^{2}E\left(\nabla,\varphi^{2}g\right)_{ab} & = & 0
\end{eqnarray}
throughout $M$, where
\begin{eqnarray}
\label{einsteintensor}
E(\nabla,\varphi^{2}g)_{ab} :=\cr
\qquad \frac{1}{2}{R(\nabla)_{acb}}^{c}+\frac{1}{2}{R(\nabla)_{bca}}^{c}-\frac{1}{2}(\varphi^{2}g_{ab})(\varphi^{-2}g^{ef}){R(\nabla)_{ecf}}^{c}
\end{eqnarray}
is the Einstein tensor defined by $\nabla_{a}$ and $\varphi^{2}g_{bc}$, whereas 
$\tilde{\nabla}$ denotes the torsion-free part of the covariant derivation 
$\nabla$, furthermore ${R(\nabla)_{abc}}^{d}$ is its Riemann tensor.
Eq.(\ref{eprelim}) can be transformed to a more familiar form via introducing 
the notation 
\begin{eqnarray}
\mathcal{T}\left(\nabla,\varphi^{2}g\right)_{ab}:= \cr
\qquad \frac{1}{4}\Biggl(\tilde{\nabla}_{a}T(\nabla)_{bg}^{g}+\tilde{\nabla}_{b}T(\nabla)_{ag}^{g}+T(\nabla)_{ga}^{h}T(\nabla)_{bh}^{g}\cr
\qquad -\frac{1}{2}(\varphi^{2}g_{ab})(\varphi^{-2}g^{ef})\left(2\tilde{\nabla}_{e}T(\nabla)_{fg}^{g}+T(\nabla)_{ge}^{h}T(\nabla)_{fh}^{g}\right)\Biggr),
\end{eqnarray}
where $T(\nabla)_{ab}^{c}$ is the torsion tensor of $\nabla$. 
Using this, Eq.(\ref{eprelim}) is equivalent to
\begin{eqnarray}
\label{e}
\tilde{\nabla}_{a}\left(\varphi^{2}g_{bc}\right) & = & 0,\cr
\varphi^{2}E(\tilde{\nabla},\varphi^{2}g)_{ab} & = & \varphi^{2}\mathcal{T}(\nabla,\varphi^{2}g)_{ab},
\end{eqnarray}
which is obtained by the well-know identity between the Riemann tensor 
of a covariant derivation and the Riemann tensor of its torsion-free part. 
The obtained field equation is nothing but an ordinary vacuum Einstein equation 
for the dimension-free metric $\varphi^{2}g_{ab}$, 
i.e.\ for the metric tensor measured in units of square Planck length $\varphi^{-2}$ in each 
spacetime point. Quite obviously, presence of matter fields will generate 
contribution to Eq.(\ref{e}) in terms of energy-momentum tensor as a source 
on the right hand side. It should be noted that whenever the torsion 
tensor $T(\nabla)_{ab}^{c}$ is not assumed to be zero a priori, it contributes to 
the energy-momentum tensor as seen from Eq.(\ref{e}).

\vspace*{2mm}
\begin{Rem}
The presented 
variational problem may be reformulated on the closed affine subspace of 
torsion-free covariant derivations, in which case the torsion tensor 
$T(\nabla)_{ab}^{c}$ automatically vanishes and thus the source term 
$\mathcal{T}(\nabla,\varphi^{2}g)_{ab}$ vanishes on the right hand side of 
Eq.(\ref{e}) along with having automatically $\tilde{\nabla}_{a}=\nabla_{a}$. 
That would mean ordinary vacuum Einstein equations for the dimension-free metric $\varphi^{2}g_{ab}$.
\label{remtorsionfree}
\end{Rem}
\vspace*{2mm}

The field equations 
Eq.(\ref{e}) may be re-expressed also in terms of the original metric 
$g_{ab}$ which is not rescaled to be dimensionless. More specifically, 
Eq.(\ref{e}) is seen to be equivalent to
\begin{eqnarray}
\label{ekg}
\tilde{D}_{a}(g_{bc}) & = & 0,\cr
E(\tilde{D},g)_{ab} & = & \mathcal{T}(\nabla,\varphi^{2}g)_{ab}\cr
 & &  + \varphi^{-1}\tilde{D}_{a}\tilde{D}_{b}\varphi + \varphi^{-1}\tilde{D}_{b}\tilde{D}_{a}\varphi \cr
 & & - 2g_{ab}\,g^{ef}\varphi^{-1}\tilde{D}_{e}\tilde{D}_{f}(\varphi) \cr
 & &  - 4\varphi^{-1}\tilde{D}_{a}(\varphi)\,\varphi^{-1}\tilde{D}_{b}(\varphi) \cr
 & & + g_{ab}\,g^{ef}\varphi^{-1}\tilde{D}_{e}(\varphi)\,\varphi^{-1}\tilde{D}_{f}(\varphi),\cr
g^{ab}\tilde{D}_{a}\tilde{D}_{b}\varphi - \frac{1}{6}\mathcal{R}(\tilde{D},g)\varphi & = & \frac{1}{6}g^{ab}\mathcal{T}(\nabla,\varphi^{2}g)_{ab}\varphi,
\end{eqnarray}
where in this case $\tilde{D}_{a}$ is a torsion-free covariant derivation over 
$L^{-1}(M){\otimes}T(M)$ such that it is metric compatible ($\tilde{D}_{a}(g_{bc})=0$), 
furthermore $E(\tilde{D},g)_{ab}$ is the Einstein tensor of $\tilde{D}_{a}$ and $g_{bc}$, 
whereas $\mathcal{R}(\tilde{D},g)$ is the Ricci scalar of 
$\tilde{D}_{a}$ and $g_{bc}$. The obtained field equation is seen to be 
nothing but the coupled conformally invariant Einstein-Klein-Gordon 
equation for $g_{ab}$ and 
$\varphi$, along with some source term coming from a possible torsion 
contribution (which may be zeroed out by means of Remark~\ref{remtorsionfree}). 
Again, when further matter fields are present, they contribute to the right 
hand side in terms of an energy-momentum tensor. The field equations 
Eq.(\ref{ekg}) are known to be conformally invariant in the conventional sense of 
Weyl rescalings.

\vspace*{2mm}
\begin{Rem}
Whenever the base manifold $M$ has a boundary $\partial{M}$ and the 
variation on the manifold boundary is allowed as 
in Remark~\ref{remarkelboundary}, the boundary field equations read as
\begin{eqnarray}
\label{eb}
\varphi^{2}g_{ab}=0\qquad \mathrm{(throughout\;}\partial{M}\mathrm{)}.
\end{eqnarray}
The field equations 
Eq.(\ref{e}) and Eq.(\ref{eb}) mean together that the dimension-free metric 
$\varphi^{2}g_{ab}$ and its Levi-Civita covariant derivation $\tilde{\nabla}_{a}$ 
obey vacuum Einstein equations with a possible additional source term 
originating from the torsion of $\nabla_{a}$. 
Furthermore, the dimension-free metric $\varphi^{2}g_{ab}$ is pressed to 
zero as approaching the boundary with a conformal scaling factor (just like 
the asymptotical behavior in the case of Friedman-Robertson-Walker 
cosmological solutions).
\end{Rem}
\vspace*{2mm}

\vspace*{2mm}
\begin{Rem}
It is worth to note that whenever the torsion is not zeroed out a priori, 
a dynamical torsion theory arises. However, due to our non-metric (Palatini-like) 
variational principle, the field equations will be slightly different than that of 
the Einstein-Cartan-Sciama-Kibble theory \cite{ortin2004}. The essential difference is: 
not the original covariant derivation $\nabla_{a}$ is compatible with the 
dimension-free metric $\varphi^{2}g_{ab}$ as in ECSK theory, but the torsion-free part of it. That is, 
if the torsion is not required to be zero a priori, the field 
equations are a simple Einstein theory for $\tilde{\nabla}_{a}$ and 
$\varphi^{2}g_{ab}$, but the torsion $T(\nabla)_{ab}^{c}$ also contributes to 
the energy-momentum tensor. In addition, one obtains the constraint equation
\begin{eqnarray}
\varphi^{-2}g^{ab}\tilde{\nabla}_{a}\mathcal{T}(\nabla,\varphi^{2}g)_{bc} = 0
\label{torsionconstraint}
\end{eqnarray}
for the torsion tensor due to the automatic vanishing of the divergence 
of the Einstein tensor because of the Bianchi identities. It is seen that Eq.(\ref{e}) along with 
Eq.(\ref{torsionconstraint}) is slightly different than that of ECSK field equations 
\cite{ortin2004}.
\end{Rem}
\vspace*{2mm}

\subsection{\bf Spinorial formulation of conformally invariant version of vacuum general relativity}
\label{spinorvacuumgr}

The proposed metric independent definition of conformal invariance 
becomes particularly useful when dealing with non-metric theories, i.e.\ with 
models in which the spacetime metric tensor is a derived quantity, not a fundamental one.

\vspace*{2mm}
\begin{Rem}
A simple example for a model in which the metric tensor is not a fundamental 
quantity can be readily given with spinorial formulation \cite{wald1984,penrose1984} 
of conformally invariant version of general relativity. In that approach, one has a spinor bundle $S(M)$ with 
two complex dimensional fibers over the real four manifold $M$. The Lagrange 
form is the spinorial representation of the conformally invariant 
Einstein-Hilbert Lagrangian (see also Eq.(\ref{eh})):
\begin{eqnarray}
\label{seh}
\LL: \cr
 \quad \Gamma\left(V(M)\,{\times}\,T^{*}(M){\otimes}V(M)\,{\times}\,T^{*}(M){\wedge}T^{*}(M){\otimes}V(M){\otimes}V^{*}(M)\right) \cr
 \qquad\qquad\qquad\qquad\qquad\qquad\qquad\qquad\quad\rightarrow \Gamma\left(\mathop{\wedge}\limits^{4} T^{*}(M)\right),\cr
 \quad ((\varphi,\epsilon_{AB},\sigma_{a}^{AA'},\chi^{B}),\,({D\varphi}_{b},{D\epsilon{}_{AB}}_{b},{D\sigma_{a}^{AA'}}_{b},{D\chi^{B}}_{b}),\,\cr
 \qquad\qquad\qquad\qquad\qquad\quad (r_{ab},{\rho_{ab}}_{AB}{}^{CD},\Pi_{ab}{}_{c}^{AA'}{}^{d}_{BB'},P_{ab}{}_{A}{}^{B}))\cr
 \quad \mapsto\;\vv(g(\sigma,\epsilon))\varphi^{2}g(\sigma,\epsilon)^{ac}\left(\sigma_{c}^{AA'}\bar{P}_{ab}{}_{A'}{}^{B'}\sigma^{b}_{AB'}+\sigma_{c}^{AA'}P_{ab}{}_{A}{}^{B}\sigma^{b}_{BA'}\right),\cr
\end{eqnarray}
where 
$V(M):=L^{-1}(M) \,\oplus\, L(M){\otimes}\mathop{\wedge}\limits^{2}S^{*}(M) \,\oplus\, T^{*}(M){\otimes}\bar{S}(M){\otimes} S(M) \,\oplus\, L^{-1}{\otimes}S(M)$.
Here, 
$g(\sigma,\epsilon)_{ab}:=\sigma_{a}^{AA'}\sigma_{b}^{BB'}\bar{\epsilon}_{A'B'}\epsilon_{AB}$ 
denotes the canonical Lorentz metric tensor generated by an 
$\epsilon_{AB}\in\Gamma\left(L(M){\otimes}\mathop{\wedge}\limits^{2}S^{*}(M)\right)$ and 
$\sigma_{a}^{AA'}\in\Gamma\left(T^{*}(M){\otimes}\bar{S}(M){\otimes} S(M)\right)$, 
furthermore $\vv(g(\sigma,\epsilon))$ denotes the canonical volume form generated 
by $g(\sigma,\epsilon)_{ab}\in\Gamma\left(L^{2}(M){\otimes}\mathop{\vee}\limits^{2}T^{*}(M)\right)$. 
In the notation, Penrose abstract indices were used according to the 
conventions of \cite{wald1984,penrose1984}. It is seen by direct substitution 
that the model defined by this Lagrange form is measure line connection invariant, 
and hence is $\D(1)$ gauge connection invariant in the sense of Section~\ref{sectionconfinv}. 
This is verified by directly observing that for any field configuration 
$\left((\varphi,\epsilon_{AB},\sigma_{a}^{AA'},\chi^{B}),\nabla_{b}\right)$ 
the Lagrangian expression
\begin{eqnarray}
 \vv(g(\sigma,\epsilon))\varphi^{2}g(\sigma,\epsilon)^{ac}\left(\sigma_{c}^{AA'}\bar{P}(\nabla)_{ab}{}_{A'}{}^{B'}\sigma^{b}_{AB'}+\sigma_{c}^{AA'}P(\nabla)_{ab}{}_{A}{}^{B}\sigma^{b}_{BA'}\right)
\end{eqnarray}
is invariant to the transformation Eq.(\ref{transf}), 
i.e.\ does not depend on the covariant derivation over the line bundle of lengths, 
where $P(\nabla)_{ab}{}_{B}{}^{D}$ is the spinorial curvature tensor of $\nabla$.
\end{Rem}
\vspace*{2mm}

\subsection{\bf Dirac kinetic term}
\label{dirackinetic}

The below example shows how the definition of conformal invariance in terms 
of connection works for the Dirac kinetic term, which is a classic example of 
nontrivial Lagrangians known to be conformally invariant in terms of Weyl rescalings.

\vspace*{2mm}
\begin{Rem}
For the definition of Dirac Lagrangian, we refer again to \cite{wald1984,penrose1984} 
for spinorial notations. One has a spinor bundle $S(M)$ with two complex 
dimensional fibers over the real four manifold $M$. The Lagrangian reads as:
\begin{eqnarray}
\label{dirac}
\LL: \cr
 \quad \Gamma\left(V(M)\,{\times}\,T^{*}(M){\otimes}V(M)\,{\times}\,T^{*}(M){\wedge}T^{*}(M){\otimes}V(M){\otimes}V^{*}(M)\right) \cr
 \qquad\qquad\qquad\qquad\qquad\qquad\qquad\qquad\quad\rightarrow \Gamma\left(\mathop{\wedge}\limits^{4} T^{*}(M)\right),\cr
 \quad ((\varphi,\epsilon_{AB},\sigma_{a}^{AA'},\chi^{B},\bar{\xi}_{C'}),\,({D\varphi}_{b},{D\epsilon{}_{AB}}_{b},{D\sigma_{a}^{AA'}}_{b},{D\chi^{B}}_{b},{D\bar{\xi}_{C'}}_{b}),\,\cr
 \qquad\qquad\qquad\qquad\qquad\quad (r_{ab},{\rho_{ab}}_{AB}{}^{CD},\Pi_{ab}{}_{c}^{AA'}{}^{d}_{BB'},P_{ab}{}_{A}{}^{B},\bar{Q}_{ab}{}_{A'}{}^{B'}))\cr
 \quad \mapsto\;\vv(g(\sigma,\epsilon))\,g(\sigma,\epsilon)^{ab}\,\sqrt{2}\,\sigma_{a}^{AA'}\Real\left(\bar{\epsilon}_{A'B'}\epsilon_{AB}\bar{\chi}^{B'}\I {D\chi^{B}}_{b}+\xi_{A}\I {D\bar{\xi}_{A'}}_{b}\right),\cr
\end{eqnarray}
where
$V(M):=L^{-1}(M)\,\oplus\,L(M){\otimes}\mathop{\wedge}\limits^{2}S^{*}(M)\,\oplus\, T^{*}(M){\otimes}\bar{S}(M){\otimes}S(M) \,\oplus\, L^{-2}(M){\otimes}S(M)\,\oplus\,L^{-1}(M){\otimes}\bar{S}^{*}(M)$.
The definition of $g(\sigma,\epsilon)_{ab}$ and $\vv(g(\sigma,\epsilon))$ is the 
same as in Section~\ref{spinorvacuumgr}. 
It is seen by direct substitution 
that the model defined by this Lagrange form is measure line connection invariant, 
and hence is $\D(1)$ gauge connection invariant in the sense of Section~\ref{sectionconfinv}. 
This is verified by directly observing that for any field configuration 
$\left((\varphi,\epsilon_{AB},\sigma_{a}^{AA'},\chi^{B},\bar{\xi}_{C'}),\nabla_{b}\right)$ 
the Lagrangian expression
\begin{eqnarray}
 \vv(g(\sigma,\epsilon))\,g(\sigma,\epsilon)^{ab}\,\sqrt{2}\,\sigma_{a}^{AA'}\Real\left(\bar{\epsilon}_{A'B'}\epsilon_{AB}\bar{\chi}^{B'}\I \nabla_{b}(\chi^{B})+\xi_{A}\I \nabla_{b}(\bar{\xi}_{A'})\right)
\end{eqnarray}
is invariant to the transformation Eq.(\ref{transf}), 
i.e.\ does not depend on the covariant derivation over the line bundle of lengths.
\end{Rem}
\vspace*{2mm}

\subsection{\bf Yang-Mills kinetic term}
\label{ymkinetic}

Our last example shows how the definition of conformal invariance in terms 
of connection works for Yang-Mills kinetic term, which is an other classic example of 
nontrivial Lagrangians known to be conformally invariant in terms of Weyl rescalings.

\vspace*{2mm}
\begin{Rem}
For the formulation of Yang-Mills Lagrangian, we postulate that our gauge 
group is a compact real Lie group. This implies that any element of a finite 
dimensional real linear representation of its Lie algebra has vanishing real 
part of its trace. The Lagrangian reads as:
\begin{eqnarray}
\label{ym}
\LL: \cr
 \quad \Gamma\left(V(M)\,{\times}\,T^{*}(M){\otimes}V(M)\,{\times}\,T^{*}(M){\wedge}T^{*}(M){\otimes}V(M){\otimes}V^{*}(M)\right) \cr
 \qquad\qquad\qquad\qquad\qquad\qquad\qquad\qquad\quad\rightarrow \Gamma\left(\mathop{\wedge}\limits^{4} T^{*}(M)\right),\cr
 \quad ((\Phi,g_{ab}),\,(D\Phi_{c},{D{g_{de}}}_{f}),\,(F_{gh},{R_{ghi}}^{j}))\cr
 \quad \mapsto\; A\,\vv(g)\,g^{ac}\,g^{bd}\,\Tr\left(\left(F_{ab}-\frac{1}{\Tr I}I\Tr F_{ab}\right)\left(F_{cd}-\frac{1}{\Tr I}I\Tr F_{cd}\right)\right)\cr
 \quad \quad +\; B\,\vv(g)\,g^{ac}\,g^{bd}\,\Imag\left(\Tr F_{ab}\right)\Imag\left(\Tr F_{cd}\right),\cr
\end{eqnarray}
where 
$V(M):=Y(M)\,\oplus\, L^{2}(M){\otimes}\mathop{\vee}\limits^{2} T^{*}(M)$. 
Here, $Y(M)$ is the vector bundle of matter fields in the Yang-Mills theory, 
possibly internally also tagged with physical dimensions in terms of tensor powers of $L(M)$. 
The symbol $\vv(g)$ means the canonical volume form generated by the 
metric tensor $g_{ab}$ as previously, and $I$ is the identity over the sections 
of $Y(M)$, whereas $\Tr$ denotes trace in terms of $Y(M){\otimes}Y^{*}(M)$. 
In the formula $A$ and $B$ are real numbers, determining the weights 
(coupling factors) of the non-abelian part and the $\mathrm{U}(1)$ part of 
the gauge group. It is seen by direct substitution 
that the model defined by this Lagrange form is measure line connection invariant, 
and hence is $\D(1)$ gauge connection invariant in the sense of Section~\ref{sectionconfinv}. 
This is verified by directly observing that for any field configuration 
$\left((\Phi,g_{ab}),\nabla_{b}\right)$ 
the Lagrangian expression
\begin{eqnarray}
 A\,\vv(g)\,g^{ac}\,g^{bd}\,\Tr\left(\left(F_{\nabla}{}_{ab}-\frac{1}{\Tr I}I\Tr F_{\nabla}{}_{ab}\right)\left(F_{\nabla}{}_{cd}-\frac{1}{\Tr I}I\Tr F_{\nabla}{}_{cd}\right)\right)\cr
 \quad \quad +\; B\,\vv(g)\,g^{ac}\,g^{bd}\,\Imag\left(\Tr F_{\nabla}{}_{ab}\right)\Imag\left(\Tr F_{\nabla}{}_{cd}\right) \cr
\end{eqnarray}
is invariant to the transformation Eq.(\ref{transf}), 
i.e.\ does not depend on the covariant derivation over the line bundle of lengths, 
where $F_{\nabla}{}_{ab}$ denotes the curvature tensor of $\nabla$ over $Y(M)$. 
This invariance property simply follows from the fact that a change of the covariant 
derivation on $L(M)$ could only give contribution through $\Real(\Tr F_{\nabla}{}_{ab})$, 
which is excluded from the Lagrangian expression by construction.
\end{Rem}

\subsection{\bf A counterexample}
\label{ymcounter}

Based on the examples presented, one could ask the question whether 
there is an counterexample, when a theory is locally $\D(1)$ gauge invariant, 
but it does not possess the extra symmetry of being independent from the choice 
of the $\D(1)$ gauge connection? The answer is affirmative: for example, a 
Yang-Mills Lagrangian based on the curvature tensor of the $\D(1)$ gauge 
connection would be conformally invariant, would be locally $\D(1)$ gauge invariant, 
but it would be an explicit function of the $\D(1)$ gauge connection, 
i.e.\ of a connection on the measure line bundle $L(M)$, and therefore 
it would not possess the property of being independent from the choice of the 
$\D(1)$ gauge connection. That is, the pertinent symmetry is not an inherent 
property of all conformally invariant Lagrangians, but it seems that the 
conformally invariant Lagrangians appearing in realistic models, such as 
the Standard Model, happen to possess this extra symmetry of being not dependent 
on the choice of a $\D(1)$ connection. With the notations as in Section~\ref{ymkinetic}, 
the pertinent non-invariant Lagrangian reads as:
\begin{eqnarray}
\label{ymcounter}
\LL: \cr
 \quad \Gamma\left(V(M)\,{\times}\,T^{*}(M){\otimes}V(M)\,{\times}\,T^{*}(M){\wedge}T^{*}(M){\otimes}V(M){\otimes}V^{*}(M)\right) \cr
 \qquad\qquad\qquad\qquad\qquad\qquad\qquad\qquad\quad\rightarrow \Gamma\left(\mathop{\wedge}\limits^{4} T^{*}(M)\right),\cr
 \quad ((\Phi,g_{ab}),\,({D\Phi}_{c},{D{g_{de}}}_{f}),\,(F_{gh},{R_{ghi}}^{j}))\cr
 \quad \mapsto\; \vv(g)\,g^{ac}\,g^{bd}\,\Real\left(\Tr F_{ab}\right)\Real\left(\Tr F_{cd}\right).\cr
\end{eqnarray}
Given a field configuration $\left((\Phi,g_{ab}),\nabla_{b}\right)$, the Lagrangian 
expression reads as
\begin{eqnarray}
 \vv(g)\,g^{ac}\,g^{bd}\,\Real\left(\Tr F_{\nabla}{}_{ab}\right)\Real\left(\Tr F_{\nabla}{}_{cd}\right).
\end{eqnarray}
Clearly, a connection on $L(M)$, possibly contained within the connection on $Y(M)$ 
will contribute to $\Real\left(\Tr F_{\nabla}{}_{ab}\right)$, and thus the 
pertinent Lagrangian is conformally invariant, but it is not invariant then to 
the choice of the connection on $L(M)$.

\section{Concluding remarks}
\label{concludingremarks}

In this paper a metric independent reformulation for the property 
of conformal invariance for classical field theories was shown. 
This was done by attaching a $\D(1)$ gauge charge (in the analogy of conformal weight) to each fundamental field. 
It was argued that the $\D(1)$ gauge charge is 
nothing but a physical dimension, and the $\D(1)$ gauge connection is 
nothing but the rule for parallel transport of measurement units between points of spacetime. 
The $\D(1)$ gauge invariance was seen to be equivalent to ordinary 
conformal invariance in terms of Weyl rescalings of the metric. It was also 
shown in addition, that conformally invariant Lagrangians turning up in physics have a 
slightly larger symmetry than that: their action functional is not only $\D(1)$ 
gauge invariant, but does not depend on the $\D(1)$ gauge connection at all.

All this was also presented by a somewhat more elegant geometrical formulation 
using measure line bundles. In that approach, the field quantities are not 
simply real valued, but each fundamental field takes its value on the tensor 
powers of line bundle of lengths. The connection on this line bundle 
corresponds to the pertinent $\D(1)$ gauge connection.

The presented metric independent variational formulation has the advantage of 
direct applicability to models in which the spacetime metric tensor is 
not a fundamental quantity, but is some function of other fundamental fields. 
(Such situation happens if the spinorial decomposition of the metric is considered 
to be the fundamental variables, rather than the metric.) 
The used metric independent formalism also reveals the discussed hidden 
symmetry of conformally invariant Lagrangians in a fairly transparent way.

\section*{Acknowledgments}

I would like to thank Tam\'as Matolcsi and to Marco Modugno from whom the 
idea of measure lines originate. I also would like to thank P\'eter V\'an, Tam\'as F\"ul\"op, 
J\'anos Krist\'of, Dezs\H{o} Varga and Vladimir Prochazka for motivations, discussion on the 
physical context and reading the manuscript. 
I would also like to thank the organizers of the ISSP workshops for 
their kind invitation.
This work was supported in part by the Momentum (`Lend\"ulet') program of the 
Hungarian Academy of Sciences under the grant number LP2013-60, and also by the 
J\'anos Bolyai Research Scholarship of the Hungarian Academy of Sciences, 
furthermore by the Hungarian Scientific Research Fund (NKFIH 123842-123959).

\bibliographystyle{unsrt}
\bibliography{confappolb}

\end{document}